\documentclass[newversion]{aa}
\usepackage{graphicx}

\begin{document}


   \title{On the Origin of the Balmer and Lyman Emission Lines}

   \author{G. Stellmacher\inst{1}
          \and
          E. Wiehr\inst{2}}

   \offprints{E. Wiehr}

   \mail{ewiehr@astrophysik.uni-goettingen.de}

   \institute{Institute d'Astrophysique (IAP), 98 bis Blvd. d'Arago, 
             75014 Paris, France
              \and
              Institut f\" ur Astrophysik der Universit\"at,
              Friedrich-Hund-Platz 1, 37077 G\"ottingen, Germany}

   \date{Received Feb. 19, 2008; accepted July 9, 2008}

\abstract
{}
{We show how the observed hydrogen Balmer and Lyman emission lines 
constrain the modeling of quiescent solar prominences.}
{We compare space observations of Lyman lines with ground-based
observations of Balmer lines for quiescent solar prominences
of comparable brightness defined by their H$\beta$ emission.}
{The effective number densities of hydrogen atoms emitting from 
the same upper level $u$ deduced from the corresponding emerging Lyman and 
Balmer line emissions show large differences that diminish with increasing 
level number and converge at the highest level numbers. Hydrogen atoms 
excited in $u = 5$ contribute 250 times less, and those in u=8 still 
contribute 65 times less to the Lyman than to the corresponding Balmer 
emission, supporting the idea of distinct spatial origin of the emissions 
of both series. This is also indicated by the line widths. The high optical 
thickness of all Lyman members allows the brightness temperature T$_b$ to be
estimated from the spectral radiance at line center, where T$_b$ is found to 
be largely independent of the upper level number, in contrast to the (known) 
behavior of the Balmer lines.}
{}

\maketitle

%
%

\section{Introduction}

The hydrogen spectrum of solar prominences is an important
source of information about the physical state and the 
structure of these cool plasma clouds embedded in a hot 
environment. The emission lines of this most abundant element 
are distributed over a wide spectral range, reaching from the 
Lyman series in the EUV to the Balmer lines in the visible and 
further series in the infrared spectral regions. Large parts 
of the EUV Lyman lines have been observed with the SUMER 
spectrograph onboard SOHO by Stellmacher et al. (2003) and by
Parenti et al. (2005a).

The complete Balmer series in the visible spectral range is much 
less frequently observed, since large dispersion spectrographs 
make simultaneous observations over the wide spectral range from 
H$\alpha$ at 6563\,\AA{} to the highest Balmer members below 388\,\AA{} 
difficult. Large parts of the Balmer series have been observed by 
Stellmacher (1969), by Yakovkin and Zel'dina (1975), and by Illing 
et al. (1975). But simultaneous observations of both the {\it whole} 
Lyman and the {\it whole} Balmer sequences do not exist to our knowledge.

However, data of both line series from different observations
may be compared using the mean values of prominences with
similar brightness. Here, the total line radiance of the (mostly
optically thin) H$\beta$ emission serves as a reasonable indicator.
Such a relation between spectroscopic data in the EUV and in
the visible spectral region may be found in De Boer, Stellmacher 
\& Wiehr (1998) and by Stellmacher, Wiehr \& Dammasch (2003) who 
simultaneously observed large parts of the Lyman lines from space 
and selected Balmer lines including H$\beta$ from ground.

%
%

\section{Comparison of observed Balmer and Lyman spectra}

To compare observed Lyman and Balmer lines, we used the equation that 
relates the spectral line radiance $E_{ul}$, i.e. the spectral radiance 
$L_{ul}$ integrated over the respective emission line, with the number 
density of atoms, $n_u$, emitting from the upper level $u$ (Uns\"old 1955): 

\begin{equation}
E_{ul} = \int L^{\lambda}_{ul} d\lambda = {h\cdot c \over 4\pi} 
\cdot {A_{ul} \over \lambda_{ul}}\cdot n_u \cdot D
\end{equation}
yielding:

\noindent\parbox{5cm}{
\begin{eqnarray*}
\log\,{n_uD \over g_u} & = &\log\,E_{ul} - \log\,{A_{ul}g_u \over \lambda_{ul}} + 
\log\,{4\pi \over hc} \\
& = & \log\,E_{ul} - C_{ul} + C 
\end{eqnarray*}}\hfill\parbox{5mm}{
\begin{equation}
\end{equation}}

\noindent
where $A_{ul}$ are the Einstein coefficients for the spontaneous
emission from upper level $u$ to lower level $l$, $g_u = 2 \cdot u^2$ 
the statistical weight of the upper level $u$; 
$C_{ul} = log(A_{ul} \cdot g_u/\lambda_{ul})$ is a constant for  
each line and given in the NBS-tables (Wiese et al. 1966), and 
$C=log(4\pi/(h\cdot c))$=16.8[cgs]. Equation\,(2) relates the 
observed spectral line radiance $E^{obs}_{ul}$ and the effective 
number of emitting atoms $n_{u}$D. For emissions from optically thin layers, 
$n_u$D is the actual number of hydrogen atoms excited in the upper level $u$ 
emitting along the line-of-sight. In the form $n_u\cdot\,D/g_u$ (in Eq. (2)), 
this quantity may be used in the Boltzmann formula. 

The effective number of emitting hydrogen atoms obtained from Lyman and 
from Balmer lines observations are compared in Fig.\,1. Its upper part 
shows means of Balmer line observations by Stellmacher (1969) and by 
Illing et al. (1975), which can be considered as characteristic of 
moderately bright prominences with emissions 
$E({\rm H}\beta)\le 2\times10^4$\,erg/(s\,cm$^2$\,ster). 
           
Substantially higher spatial resolution is achieved in the detailed 
photometric data of H$\alpha$ and H$\beta$ (u = 3; 4) by Stellmacher 
\& Wiehr (1994). The results from observations by Yakovkin \& Zel'dina
(1975) are not entered in Fig. 1 since comparable Lyman observations 
do not exist, and they represent rare prominences with very bright 
emissions up to $E({\rm H}\beta)\le 16.5\times10^4$\,erg/(s\,cm$^2$\,ster),
like those discussed by Stellmacher \& Wiehr (2005).

For the faint and moderately bright prominences with $E({\rm
  H}\beta)\le\,2\times\,10^4$\,erg/(s\,cm$^2$\,ster), EUV observations
with the SUMER instrument onboard SOHO have been taken by De\,Boer,
Stelllmacher \& Wiehr (1998) and by Stellmacher, Wiehr \& Dammasch
(2003) simultaneously with ground-based data. These Lyman data are
shown in the lower part of Fig. 1 together with EUV observations by
Heinzel et al.  (2001) and by Parenti, Vial \& Lemaire (2005a).

The differences in the abscissa values for equal upper level $u$ of 
the Balmer and the Lyman lines arise from the different Einstein 
coefficients $A_{ul}$. The striking differences in the ordinate values 
of the Lyman and the Balmer lines reflect the largely different number 
of effectively emitting atoms. Emerging Lyman lines stem from up to 250 
times less emitters than the corresponding Balmer lines from the same 
upper level u.
                         
%

   \begin{figure}[ht]     
   \includegraphics[width=8.8cm]{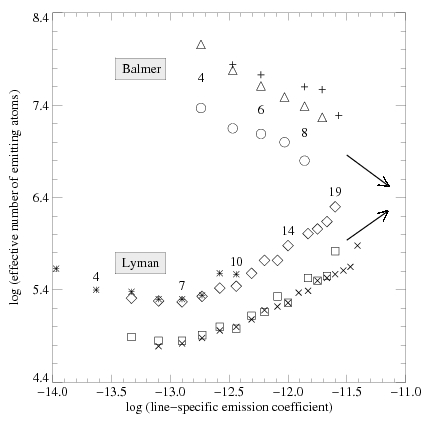}
   \caption{Effective number of atoms, log $E^{obs}_{ul}-C_{ul}-+C$, emitting 
from the upper level $u$ to the lower level l, versus the line-specific 
emission coefficient, $C_{ul}=log(A_{ul}\cdot g_{ul}/\lambda_{ul})$, from
observations of the Balmer series by Stellmacher (1969; trangles for 
faint and circles for brighet prominences) and by Illing, Landman \&
Mickey (1975; crosses) in comparison with observations of the Lyman 
series by Stellmacher, Wiehr \& Dammasch (2003; rhombs=42L, squares=70-H),  
by Parenti, Vial \& Lemaire (2005; X), and by Heinzel et al. (2001; 
asterisks); indices = upper level numbers u.}\label{FigGam}
    \end{figure}

%
%

Figure 1 shows a steady decrease for the Balmer lines in the ordinate
values $\log\,E_{ul}-C_{ul}+C$ with increasing upper level $u$, and
this reflects the decreasing transition probabilities. The Lyman lines
show an opposite behavior of increasing ordinate values with
increasing upper level for $u\ge8$, approaching the values for the
Balmer lines. Only the centrally depressed Lyman lines with $u\le7$
show a slight decrease in the numbers of effectively emitting atoms
for increasing u.

For an interpretation of the different behavior of Balmer and Lyman
lines, we follow the NLTE calculations by Gouttebroze, Heinzel \& Vial
(1993).  The analysis of the Balmer lines by Stellmacher (1969) shows
that the populations of the upper levels $u>4$ follow, relative to
each other, a smooth distribution with an equilibrium temperature of
$T\approx6000$\,K.  This value is close to the kinetic temperature,
$T_{kin} (\approx T_e)$, obtained from the line widths and is conform
to the departure coefficients $b_u=1$ for $u>4$ calculated by
Gouttebroze et al. (1993).

For the Lyman lines, these authors obtain a total optical prominence
thickness equivalent to $\tau({\rm Ly}_{22\rightarrow 1})=\,230$ with
'conventional' electron temperature $T_e=8000$\,K. For higher
temperatures of a prominence corona transition atmosphere,
$T=10\,000$\,K ($T=15\,000$\,K), their models still lead to values of
$\tau$(Ly$_{22\rightarrow 1})=48$ (2.5), and even Ly$_{8\rightarrow
  1}=1000$ (52).

Emission lines with such high $\tau$ values will be limited to the
outermost (hotter) prominence regions. A rough estimate shows that the
geometric extention of the effectively emitting layers (which produce
the emerging Lyman emission) is rather small as compared to that of
the (optically thin) Balmer lines, which amounts to the whole
prominence atmosphere. In general, the geometric thickness H of an
emitting layer can be estimated from
\begin{equation}        
\tau_{u,1} = { e^2\cdot \sqrt(\pi) \over m_e \cdot c^2} 
\cdot {\lambda \over \Delta\lambda}
\cdot \lambda\cdot f \cdot n_1 \cdot H 
\end{equation} 
with $n_1\approx 0.5 \cdot n_e = 2 \times 10^{10} cm^{-3}$ and
$\Delta\lambda/\lambda = 1.2 \times 10^{-4}$ (as observed), one
obtains for Ly$_{8\rightarrow 1}$ a geometric thickness of only $H =
30$\,km for an assumed high value of $\tau$(Ly$_{8\rightarrow 1})=20.$

With increasing level number u, the total thickness decreases toward 
that of the Lyman continuum, for which Parenti, Vial \& Lemaire (2005b) 
still deduce $\tau({\rm Ly}_{\infty\rightarrow 1})\approx 6.5$, and the emissions 
will originate more and more in deeper (cooler) layers, thus approaching
the main regions of Balmer line emissions. This readily explains both 
the 'bending' of the Lyman curves in Fig. 1 and their approach to the 
values of the optically thin ($u>4$) Balmer lines. Upper levels $u\ge12$ 
will largely be in equilibrium with the free electrons since their energy 
difference from the ionization limit amounts to only $\Delta E < 0.1$\,eV.
                      
%

   \begin{figure*}[htb]     
   \hspace{0mm}\includegraphics[width=18.5cm]{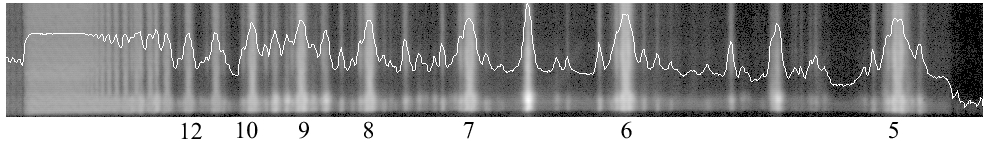}
   \caption{SUMER spectrum of the prominence observed on July\,8, 2000 by 
Stellmacher, Wiehr \& Dammasch (2003), showing the Lyman emissions 
(u$\ge5)$ in the wavelength range 910\,\AA$<\lambda<$947\,\AA{} along
the 120'' slit, which covers 20'' of the upper chromopshere (lower 
image border), 30'' regions below the prominence, and 70'' of the 
prominence body. The white line gives the spectral radiance distribution 
averaged over the prominence body in a logarithmic scale. The jump 
between u=10 and u=11 is due to the different detector coating (KBr 
for larger $\lambda$). Note the largely constant spectral radiance of 
the Lyman line centers.}
            \label{FigLev}
    \end{figure*}

%
%

\section{Spectral radiance at the Lyman line centers}

The spectral radiances at line center $L_{u1}(\lambda_0)$ are largely 
independent of u. This can be seen in the SUMER spectrum (Fig. 2)
showing the Lyman emissions $(u \ge 5)$ in the wavelength range
910\,\AA$ <\lambda<$ 947\,\AA{} along the 120'' slit, which covers 70'' 
of the prominence body. Superposed is the mean spectral radiance 
distribution $L_{u1}(\lambda)$ over the prominence body in a 
logarithmic scale.

The almost constant central line radiances $L_{u1}(\lambda_0)$ indicate 
largely equal brightness temperatures $T_b$, which may be derived via the 
Planck function. For the optically thick Lyman lines ($\tau(Ly)>>1$, 
hence emitted from the prominence periphery), the spectral radiance at 
line center $L_{u1}(\lambda_0)$ becomes equal to the source function 
$S_{u1}$, which we set equal to the Planck function $B_{u1}(T_b).$

In Table 1 we give the deduced $T_b$, together with the observed 
spectral radiance at the Lyman line centers L$_{u1}(\lambda_0)$ and
the total spectral line radiance $E_{ul}=\int L_{ul}^{\lambda}$ 
for the brightest (42-L) and the faintest (70-H) prominence regions 
marked in Fig. 1 as rhombs and squares. The mean values of, 
respectively, $T_b=4989$\,K and $T_b=4822$\,K (neglecting the 
centrally most reversed line $u = 5$) show a small internal scatter 
of only $\pm20$\,K. The radiometric accuracy of 15\% for detector\,A 
in the corresponding $\lambda$ regime (Sch\"uhle et al. 2000) 
introduces an uncertainty of $\Delta T_b= 20$\,K.

The $T_b$ values deduced for the Lyman lines are more than 1000\,K 
higher than the $T_{ex}$ deduced from the Balmer lines, for which 
Stellmacher (1969) finds 3250\,K$<T_{ex}<3840$\,K in similarly 
bright prominences. Even the maximum value $T_{ex}=3900$\,K, 
occasionally found for the (optically thick) H$\alpha$ line in the 
brightest prominences by Stellmacher \& Wiehr (2005) is still 
$\approx1000$\,K below the $T_b$ found for the Lyman lines. This 
indicates that the peripheral prominence regions, which emit the
Lyman lines, are significantly hotter than the prominence cores,
which emit the Balmer lines.
%
%

   \begin{table}[t]
      \caption[]{Upper level u, wavelangth $\lambda_0$[\AA]{}, spectral line radiance 
$E^{obs}_{ul}$[mW/(s\,m$^2$\,ster)], spectral line radiance at line center 
$L_{ul}(\lambda_0)$[mW/(s\,m$^2$\,ster\,\AA)]{}, corresponding brightness temperature
$T_b$[K] for the brightest (42-L), and the faintest (70-H) prominence region observed by 
Stellmacher, Wiehr \& Dammasch (2003) as in Fig.\,1.}
      \[
      \begin{array}{rccccccccccc}
      \hline
      \noalign{\smallskip}
  u & \lambda_0 &&&           & {\rm brightest} &      &&&            &{\rm faintest}  &     \\ \\

    &           &&&E^{obs}_{ul}& L_{il}(\lambda_0& T_b  &&& E^{obs}_{ul}&L_{il}(\lambda_0 & T_b \\
\hline \\
  5 &   949.8   &&&    69.1   &      0.057      & 4899 &&&    25.9   &      0.022      & 4752 \\
  6 &   937.8   &&&    36.8   &      0.077      & 4999 &&&    14.1   &      0.031      & 4855 \\
  7 &   930.8   &&&    23.3   &      0.056      & 4979 &&&     9.0   &      0.023      & 4841 \\ 
  8 &   926.2   &&&    18.0   &      0.051      & 4985 &&&     6.9   &      0.020      & 4840 \\
  9 &   923.2   &&&    15.7   &      0.045      & 4979 &&&     6.0   &      0.016      & 4820 \\
 10 &   921.0   &&&    12.1   &      0.039      & 4966 &&&     4.2   &      0.014      & 4809 \\ 
 11 &   919.4   &&&    12.4   &      0.040      & 4977 &&&     4.2   &      0.015      & 4827 \\
 12 &   918.1   &&&    13.3   &      0.042      & 4991 &&&     3.6   &      0.012      & 4799 \\
 13 &   917.2   &&&    10.2   &      0.037      & 4975 &&&     4.2   &      0.016      & 4864 \\ 
 14 &   916.4   &&&    11.9   &      0.045      & 5009 &&&     2.9   &      0.011      & 4794 \\ 
 15 &   915.8   &&&      -    &        -        &   -  &&&      -    &        -        &   -  \\ 
 16 &   915.3   &&&    11.0   &      0.037      & 4983 &&&     3.6   &      0.014      & 4835 \\
 17 &   914.9   &&&    10.4   &      0.037      & 4985 &&&     7.9   &      0.012      & 4814 \\
 18 &   914.6   &&&    10.3   &      0.034      & 4973 &&&     2.6   &      0.012      & 4815 \\ 
 19 &   914.3   &&&    13.1   &      0.033      & 4970 &&&     4.2   &      0.011      & 4803 \\ 
 \hline

         \end{array}
         \]
   \end{table}

%
%

\subsection{Influence of filling}
          
Since most prominences are formed by numerous tiny 'threads' (e.g. 
Lin et al. 2005), the actual spectral radiances at line center can 
be expected to be higher than the observed ones. The coronal material 
inbetween will not emit Lyman lines. Realistic filling factors are 
still unknown. If we assume a filling of 50\%, the actual spectral 
radiances would be two times, for a 1\% filling 100 times greater 
than those in Table 1. The influence of filling on the eventually 
deduced $T_b$ values is, however, rather weak, since even our brightest 
Lyman line at $\lambda=937.8$\,\AA{} yields for a 50\% (10\%) filling 
an increase of $T_b$ by only 110\,K (400\,K). 

Such a weak influence of filling also seems to be indicated when comparing 
observations of different prominences. In particular, emissions deduced 
from two-dimensional imaging at much higher spatial resolution (Stellmacher 
\& Wiehr 1999; Stellmacher \& Wiehr 2000) largely follow the same relations 
as from spectra of much lower spatial resolution. 

\section{Line widths}                    

High temperatures for the Lyman emitting regions are also indicated
from the line widths. $\Delta\lambda/\lambda = 0.4 \times 10^{-4}$
observed for the unsaturated higher Balmer lines by Stellmacher (1969)
and by Stellmacher \& Wiehr (1994) correspond for purely thermal
broadening to $T_{kin}\approx8700$\,K, repectively for a mean
non-thermal broadening of 4\,km/s to $T_{kin}\approx7500$\,K.

In contrast, the widths $\Delta\lambda/\lambda=1.1\times 10^4$
measured by Parenti, Vial \& Lemaire (2005a) for the higher Lyman
lines with $u>16$, correspond for purely thermal broadening to an
upper limit (since optically thick) of $T^{max}_{kin}\approx
66\,000$\,K. For the hydrogen formation temperature of 16\,000\,K, a
nonthermal broadening of 28.5\,km/s is required, in accordance with
Stellmacher, Wiehr \& Dammasch (2003; Figs. 16 and 17) and with
Parenti \& Vial (2007). However, the typical kinetic temperature of
7500\,K, deduced from the Balmer lines for the cool prominence body,
would lead to a high non-thermal broadening of 31\,km/s.

Slightly broader Lyman lines with $\Delta\lambda/\lambda=1.25\times10^{-4}$ 
were measured by Stellmacher, Wiehr \& Dammasch (2003). They discuss that 
the actual values may be smaller, since the intrinsic SUMER profile seems 
to be underestimated. The EUV spectrograph is particularly adapted to broad 
emission lines from the hot corona rather than to narrow lines from cool 
prominences. Regardless of that uncertainty, the broadening of the Lyman 
lines yields much higher temperatures T$_{kin}$ than that of the Balmer lines, 
in accordance with the difference in the brightness temperatures T$_b$.

\section{Conclusions}

The comparison of observed Lyman and Balmer emissions from faint
through 'medium bright' quiescent prominences, characterized by
$E({\rm H}\beta) < 2 \times 10^4$\,erg/(s\,cm$^2$\,ster), shows that
the inner regions with $T_{kin} \approx 7500$\,K that emit the
optically thin Balmer lines are not those with much higher $T_{kin}$
that emit the Lyman lines. Even different members of the Lyman series
will not originate in the same prominence volume. The very high $\tau$
values of the first Lyman members will limit their emerging emission
to the outermost prominence periphery where excitation and ionization
are highest.

With decreasing $\tau\approx\lambda\cdot\,f$, higher Lyman members 
will originate more and more in the deeper layers. A realistic modeling
of the Lyman and the Balmer lines will have to consider strong gradients 
between the cool prominence body and its hot periphery for temperature 
and (or) non-thermal velocities besides the strong departures from LTE.

Heinzel et al. (2001) propose that the reversed profiles of the stronger 
Lyman lines will be related to the orientation of the line-of-sight (LOS) 
with respect to themagnetic field lines: emissions viewed across the field 
lines are expected to show strong rever- sals. The calculations by Loucif 
\& Magnan (1982) can be useful for profile modeling. They treat the 
transfer problem of emerging reversed lines taking spatially correlated 
velocity fields into account and applying the method of addition of layers.  

From our various observations, we do not find systematic differences between 
prominences viewed under different aspect angles, but instead we find the 
largely unique emission relations displayed in Fig.\,1. Observations at higher 
spatial resolution in- cluding the magnetic field might help for adapting more 
refined models. 

%
%

\begin{acknowledgements}
We thank Drs. F.\,Hessman and I.\,E.\,Dammasch and for helpful discussions; 
B.\,Bovelet kindly performed the graphics in Fig.\,1.     
\end{acknowledgements}

%
%

%
%


\begin{thebibliography}{}

\bibitem{} de\,Boer, C. R., Stellmacher, G., \& Wiehr, E. 1998, A\&A, 334, 280 

\bibitem{}Gouttebroze, P., Heinzel P., \& Vial J.-C. 1993, A\&AS, 99, 513 

\bibitem{} Heinzel, P., Schmieder, B., Vial, J.-C., \& Kotrc, P. 2001, A\&A, 370, 281 

\bibitem{} Illing, R. M. E., Landman, D.A., \& Mickey, D. L. 1975, Sol. Phys., 45, 339 

\bibitem{} Loucif, M. L., \& Magnan, C. 1982, A\&A, 112, 287 

\bibitem{} Lin, Y., Engvold, O., Rouppe van der Voort, L.,Wijk, J. E., \& Berger, T. E. 2005 Sol. Phys., 226, 239 

\bibitem{} Parenti, S., Vial, J.-C., \& Lemaire, P. 2005a, A\&A, 443, 679 
 
\bibitem{} Parenti, S., Vial, J.-C., \& Lemaire, P. 2005b, A\&A, 443, 685 

\bibitem{} Parenti, S., \& Vial, J.-C. 2007 ApJ, 469, 1109 

\bibitem{} Sch\"uhle, U., Curdt, W., Hollandt, J., et al. 2000, Appl. Opt., 36, 6416 

\bibitem{} Stellmacher, G. 1969, A\&A, 1, 62 

\bibitem{} Stellmacher, G., \& Wiehr, E. 1994, A\&A, 290, 655 

\bibitem{} Stellmacher, G., \& Wiehr, E. 2005, A\&A, 431, 1069 

\bibitem{} Stellmacher, G., Wiehr, E., \& Dammasch, I. E. 2003, Sol. Phys., 217, 133 

\bibitem{} Uns\"old, A. 1955, Physik der Sternatmosph\"aren, 2.Aufl. (Berlin: Springer) 

\bibitem{} Wiese, W. L., Smith, M.W., Glennon, B. M. 1966, Atomic Transition Probabilities, 
Vol. I, H through Ne, Nat. Stand. Ref. Data Ser., Nat. Bur. Stand. (US) 4, 
U.S.-Gov.Washington DC: Printing Office 

\bibitem{} Yakovkin, N. A., \& Zel'dina, M.Yu. 1975, Sol. Phys., 45, 319



\end{thebibliography}
\end{document}